\documentclass[preprint,aps,floats,nofootinbib]{revtex4}
\input epsf.tex
\def\DESepsf(#1 width #2){\epsfxsize=#2 \epsfbox{#1}}
\usepackage{graphicx}
\usepackage{psfig}

\def\thebibliography#1{\centerline{\bf REFERENCES}
  \list{[\arabic{enumi}]}{\settowidth\labelwidth{[#1]}\leftmargin
  \labelwidth\advance\leftmargin\labelsep\usecounter{enumi}}
\def\newblock{\hskip .11em plus .33em minus -.07em}\sloppy
  \clubpenalty4000\widowpenalty4000\sfcode`\.=1000\relax}

\begin{document}

\draft

\vspace*{0.5cm}

\title{Exclusive $B\to M \nu \bar{\nu}~(M= \pi, K, \rho, K^*)$ Decays \\
and  Leptophobic $Z^\prime$ Model}

\author{
Jong~Hun~Jeon$^1$\footnote{jhjeon@cskim.yonsei.ac.kr}, ~
C.~S.~Kim$^1$\footnote{cskim@yonsei.ac.kr}, ~
Jake~Lee$^2$\footnote{jilee@cskim.yonsei.ac.kr} ~
and ~
Chaehyun~Yu$^3$\footnote{chyu@korea.ac.kr}}

\affiliation{
$^1$ Department of Physics, Yonsei University,
Seoul 120-479, Korea \\
$^2$ Department of Physics, Sogang University,
Seoul 100-611, Korea \\
$^3$ Department of Physics, Korea University,
Seoul 136-701, Korea
~\vspace{1cm}
}
\date{\today}

\begin{abstract}
\noindent
We consider the exclusive flavor changing neutral current processes
$B \rightarrow M \nu \bar{\nu}~(M= \pi, K, \rho, K^*)$
in the leptophobic $Z^\prime$ model,
in which the charged leptons do not couple to the extra $Z^\prime$ boson.
We find that these exclusive modes are very effective
to constrain the leptophobic $Z^\prime$ model.
In the leptophobic $Z^\prime$ model, additional right-handed neutrinos
are introduced and they can contribute to the missing energy signal
in $B\to M + E\hspace{-0.25cm}/$ decays.
Through the explicit calculations, we obtain
quite stringent bounds on the model parameters,
$|U_{sb}^{Z^\prime}| \leq 0.29$ and
$|U_{db}^{Z^\prime}| \leq 0.61$,
from the already existing experimental data.
We also briefly discuss an interesting
subject of massive right-handed neutrinos, which might be connected with
the dark matter problem.

\end{abstract}
\maketitle

%%%cover page end

%%%%%%%%%%%%%%%%%%%%%%%%%%%%%%%%%%%%%%%%%%
%%%%%%%%%%%%%%%%%%%%%%%%%%%%%%%%%%%%%%%%%%
\section{introduction}
%%%%%%%%%%%%%%%%%%%%%%%%%%%%%%%%%%%%%%%%%%
%%%%%%%%%%%%%%%%%%%%%%%%%%%%%%%%%%%%%%%%%%

Flavor changing neutral current (FCNC) processes have gained continuous
interests as powerful probes of new physics beyond the standard model (SM)
as well as  stringent tests within the SM.
Since FCNC processes are only generated at the loop level within the SM,
they are suppressed by the loop momentum and off-diagonal
Cabibbo-Kobayashi-Maskawa (CKM) matrix elements.
Therefore, if FCNC exists in the tree level or exists with an enhancement
factor in the coupling without  severe suppression, it would be very
sensitive to new physics effects beyond the SM.
$B$-factory experiments such as Belle and BaBar
have given various opportunities to test those FCNC processes.
Among the various terms in the effective Hamiltonian responsible
for the decay of  $b$ quark, the FCNCs are represented
by QCD penguin and electroweak (EW) penguin operators.
There are plenty of QCD or EW penguin dominant processes
in non-leptonic and semi-leptonic $B$ decays.
As an example, recent experimental data for QCD penguin dominant $B\to K \pi$ decays
appear to be very interesting:
Branching ratios and direct and mixing-induced CP asymmetries
could not be consistently explained within the current SM frameworks,
for example, QCD factorization (QCDF) \cite{Beneke:1999br},
perturbative QCD (PQCD) \cite{Keum:2000wi},
and soft-collinear effective theory (SCET) models \cite{Bauer:2005kd}.
Although it has been claimed in Ref.~\cite{Li:2005kt}
that all the current data except for the mixing-induced CP asymmetry
could be accommodated within the PQCD scheme by considering
the next-leading-order corrections,
there is still room
for new physics or new mechanism beyond the SM, especially,
in the EW penguin sector
\cite{Mishima:2004um,Arnowitt:2005qz,Barger:2003hg}
to explain all the experimental data simultaneously.

The EW penguin contributions have been thoroughly studied in the SM and beyond, too:
While the EW penguin effect with the photon emission in
$b\to s \gamma$ processes has been precisely determined,  and already put
stringent bounds on the relevant new physics scenarios,
experimental measurements for the $Z$-mediated EW penguin effect
are now under way at $B$ factories through
$B\to X \ell \ell$ and $B\to X \nu \bar{\nu}$ decay processes.
Although it is experimentally very difficult to measure,
$ B\to X \nu \bar{\nu}$ decay is an extremely good channel for the study of
$Z$-mediated EW penguin contribution within the SM and beyond
\cite{Grossman:1995gt,Buchalla:2000sk,Aliev:1997se,Aliev:2001in,Bobeth:2001jm}.

In many new physics scenarios, an extra $U(1)$ gauge boson is usually
introduced after the high energy symmetry breaking or as a low energy effective theory
\cite{Hewett:1988xc,Leike:1998wr,Chay:1998hd}.
In general, the extra gauge boson, commonly named as  $Z^\prime$ boson,
has a possibility to induce the FCNCs.
In the present work, we focus on the $E_6$-based
leptophobic $Z^\prime$ model \cite{Leroux:2001fx} among
various possibilities with the extra $U(1)$ gauge boson extension
such as the string-inspired grand unified theories (GUTs)
\cite{Hewett:1988xc,Cvetic:1996mf},
dynamical symmetry breaking models \cite{Buchalla:1995dp},
string inspired models \cite{Cleaver:1998gc},
extra dimension models \cite{Masip:1999mk},
and little higgs models \cite{littlehiggs}.
It is well known that the extra neutral gauge boson can be leptophobic
by introducing the kinetic mixing term in the $E_6$-based model.
Originally, the leptophobic  $Z^\prime$ model was
introduced to explain the $R_b$-$R_c$ puzzle at LEP \cite{:1995fx}
and anomalous high-$E_T$ jet cross section at CDF \cite{Abe:1996wy}.
Although its original motivation has been disappeared,
the leptophobic $Z^\prime$ model
certainly remains as a viable scenario beyond the SM.
The main experimental constraints on the (non-leptophobic) $Z^\prime$ boson arise from
the electroweak precision measurements at LEP
or direct searches via Drell-Yan processes at Tevatron.
For the leptophobic $Z^\prime $ boson, however,
we have now only one experimental constraint from
the dijet experiments at D0
which exclude the mass range for
$365~{\rm GeV} \leq M_{Z^\prime} \leq 615~{\rm GeV}$ \cite{Abbott:1997dr}.
For the future experiments, while
the CERN Large Hadron Collider (LHC) could serve as a probe
of the leptophobic $Z^\prime$ boson via $pp \to \gamma + E\hspace{-0.3cm}/
\hspace{0.07cm}$,
the future $e^- e^+$ linear collider (ILC)
 may have no role as far as leptophobia is concerned.
In this regard, we note that
$B \to M + E \hspace{-0.3cm} /\hspace{0.1cm}$
 decays, where $M$ is a
pseudoscalar or a vector boson, will be a good channel
for the probe of the leptophobic $Z^\prime$ model.

In this paper, we investigate
$B \rightarrow M \nu \bar{\nu}~ (M= \pi, K, \rho, K^*)$ decays
in the leptophobic $Z^\prime$ model as a possible candidate
of  new physics in the EW penguin sector.
First, we briefly introduce the leptophobic $Z^\prime$ model
based on the $E_6$ grand unified theory (GUT) in Sec.~\ref{section2}.
Sec.~\ref{section3} deals with the effective Hamiltonian and decay rates in detail.
In Sec.~\ref{section4} we present numerical analysis and discuss its implications.
Concluding remarks are also in Sec.~\ref{section4}.

%%%%%%%%%%%%%%%%%%%%%%%%%%%%%%%%%%%%%%%%%%
%%%%%%%%%%%%%%%%%%%%%%%%%%%%%%%%%%%%%%%%%%
\section{Leptophobic $Z^\prime$ model and FCNC}
\label{section2}
%%%%%%%%%%%%%%%%%%%%%%%%%%%%%%%%%%%%%%%%%%
%%%%%%%%%%%%%%%%%%%%%%%%%%%%%%%%%%%%%%%%%%

First, we briefly review the leptophobic $Z^\prime$ model.
In  GUT or string-inspired point of view,
the $E_6$ model is a very plausible extension of the SM
\cite{Rizzo:1998ut}.
It is natural that  a $U(1)^\prime$ gauge group remains as
a low energy effective theory after the symmetry breaking of the $E_6$ group.
We assume that  the $E_6$ group is broken through the following breaking chain
\begin{eqnarray}
&&E_6 \to SO(10) \times U(1)_\psi \nonumber \\
&&\phantom{E_6 }  \to SU(5) \times U(1)_\chi \times U(1)_\psi \nonumber \\
&&\phantom{E_6 }  \to SU(2)_L \times U(1) \times U(1)^\prime ,
\end{eqnarray}
where $U(1)^\prime$ is a linear combination of two additional
$U(1)$ gauge groups with
$$
Q^\prime = Q_\psi \cos \theta - Q_\chi \sin \theta,
$$
where $\theta $ is the familiar $E_6$ mixing angle.

The most general Lagrangian, which is  invariant under the SM gauge group
with an extra $U(1)^\prime$,
allows the kinetic mixing term
$\displaystyle {\cal L}_{\rm mixing} = - \frac{\sin \chi}{2} \tilde{B}_{\mu\nu}
\tilde{Z}^{\prime ~\mu\nu} $ between the $U(1)$ and $U(1)^\prime$ gauge boson fields.
This off-diagonal term can be removed by the non-unitarity transformation
\begin{equation}
\tilde{B}_\mu = B_\mu - \tan \chi Z_\mu^\prime,
~~\tilde{Z}_\mu^\prime = \frac{Z_\mu^\prime}{\cos \chi},
\end{equation}
which leads to the possibility of leptophobia of the physical $Z^\prime$ gauge boson
with the $E_6$ mixing.
Once all the couplings are GUT normalized,
the interaction Lagrangian of fermion fields and $Z^\prime$ gauge
boson can be written as
\begin{equation}
{\cal L}_{\rm int} = - \lambda \frac{g_2}{\cos \theta_W}
\sqrt{\frac{5 \sin^2 \theta_W}{3}}
\bar{\psi} \gamma^\mu \left( Q^\prime + \sqrt{\frac{3}{5}}\delta Y_{SM} \right)
\psi Z_\mu^\prime ,
\end{equation}
where the ratio of gauge couplings $\lambda = g_{Q^\prime}/g_Y$,
and $\delta=-\tan \chi/\lambda$ \cite{Rizzo:1998ut}.
The general fermion-$Z^\prime$ couplings depend on two free parameters,
$\tan \theta$ and $\delta$, effectively \cite{Babu:1996vt}.
The $Z^\prime$ boson can be
leptophobic when $(Q^\prime + \sqrt{\frac{3}{5}}\delta Y_{SM})=0$
for $L$ and $e^c$ simultaneously.
$L$ and $e^c$ are the lepton fields in the {\bf 27} representation of
$E_6$, as shown in Table~\ref{table:embedding}.
In the conventional embedding,
the $Z^\prime$ boson can be made leptophobic
with $\delta=-1/3$ and $\tan \theta = \sqrt{3/5}$.
In other embedded schemes, where the quantum numbers
between $L$ and $H$ (or $e^c$, $\nu^c$, and $S^c$) are switched,
one can find suitable values for $\delta$ and
$\tan \theta$ which make the $Z^\prime$ boson leptophobic \cite{Leroux:2001fx}.

%%%%%%%%%%%%%%%%%%%%%%%%%%%%%%%%%%%%%%%%%%
%%%%%%%%%%%%%%%%%%%%%%%%%%%%%%%%%%%%%%%%%%
% Table I
\begin{table}
\caption{ Charges of fermions contained in the {\bf 27}
representation of $E_6$ within the conventional embedding \cite{Rizzo:1998ut}.
 }
\smallskip
\label{table:embedding}
\begin{tabular}{c|cccc}
\hline
Particle &~~ $SU(3)_c$~~         &~~ $Y$ ~~
         &~~ $2\sqrt{6}Q_\psi$~~ &~~ $2\sqrt{10}Q_\chi$~~  \\
\hline
$Q=(u,d)^T$       & ${\bf 3}$       & $\phantom{-} 1/6$
                  & $\phantom{-} 1$ & $-1$ \\
$L=(\nu,e)^T$     & ${\bf 1}$       & $-1/2$
                  & $\phantom{-} 1$ & $\phantom{-} 3$ \\
$u^c$             & ${\bf \bar{3}}$ & $-2/3$
                  & $\phantom{-} 1$ & $-1$ \\
$d^c$             & ${\bf \bar{3}}$ & $\phantom{-} 1/3$
                  & $\phantom{-} 1$ & $\phantom{-} 3$ \\
$e^c$             & ${\bf 1}$       & $\phantom{-} 1\phantom{/6}  $
                  & $\phantom{-} 1$ & $-1$ \\
$\nu^c$           & ${\bf 1}$       & $\phantom{-} 0\phantom{/6}
                  $ & $\phantom{-} 1$ & $-5$ \\
$H=(N,E)^T$       & ${\bf 1}$       & $-1/2$
                  & $-2$            & $-2$ \\
$H^c=(N^c,E^c)^T$ & ${\bf 1}$       & $\phantom{-} 1/2$
                  & $-2$ & $\phantom{-} 2$ \\
$h$               & ${\bf 3}$       & $-1/3$
                  & $-2$ & $\phantom{-} 2$ \\
$h^c$             & ${\bf \bar{3}}$ & $\phantom{-} 1/3$
                  & $-2$            & $-2$ \\
$S^c$             & ${\bf 1}$       & $\phantom{-} 0\phantom{/6}  $
                  & $\phantom{-} 4$ & $\phantom{-} 0$ \\
\hline
\end{tabular}
\end{table}
%%%%%%%%%%%%%%%%%%%%%%%%%%%%%%%%%%%%%%%%%%
%%%%%%%%%%%%%%%%%%%%%%%%%%%%%%%%%%%%%%%%%%

In the leptophbic $Z^\prime$ model, FCNCs can arise
through the mixing between the ordinary SM fermions and exotic fermions.
Since the mixing of the left-handed fermions may lead to the relatively
large $Z$-mediated FCNCs,
we only allow the mixing of the right-handed fermions.
That is,  the mixing between $d_R (s_R, b_R)$ and $h_R$ can induce
the FCNCs when their $Z^\prime$ charges are different \cite{Leroux:2001fx}.
After integrating out degrees of freedom of heavy exotic fermions
and gauge bosons,
the FCNC Lagrangian for the $b\to q (q=s,d)$ transition can be parameterized as
\begin{equation}
{\cal L}_{\rm FCNC}^{Z^\prime} = - \frac{g_2}{2 \cos \theta_W}
U_{qb}^{Z^\prime} \bar{q}_R \gamma^\mu b_R Z_\mu^\prime ,
\end{equation}
where all the theoretical uncertainties including the mixing parameters
are absorbed into the coupling $U_{qb}^{Z^\prime}$.
Then, the experimental constraints on $U_{qb}^{Z^\prime}$
may be obtained by considering related non-leptonic $B$ decays
such as $B\to K \pi$ decays for which one may also find
a clue to a solution of the {\it recently--popular} $B \to K \pi$ puzzle.
But, it is not easy to obtain reliable bounds on
$U_{qb}^{Z^\prime}$ because the non-leptonic $B$ decays severely depend
on unknown strong phases  \cite{Kim:2006} and, furthermore, the EW penguin amplitude
is sub-dominant in most non-leptonic decays.

It is obvious that if new physics effects are mainly through the EW penguin contributions,
the EW penguin dominant process can give the
most powerful constraints,
and it is natural to consider the processes
such as $b \to s \gamma, b\to s(d) \ell \ell, b \to s(d) \nu \bar{\nu}$
as appropriate probes of the relevant new physics.
In the most new physics scenarios such as the minimal supersymmetric
standard model (MSSM),
model parameters are connected to many different observables, and
therefore, in order to get consistent constraints on the model
parameters we have to consider all the relevant processes simultaneously.
As we will see later, however,
one can consider only $ b \to s(d) \nu \bar{\nu}$ decays
for the leptophobic $Z^\prime$ model and almost free from other
experimental constraints.

%%%%%%%%%%%%%%%%%%%%%%%%%%%%%%%%%%%%%%%%%%
%%%%%%%%%%%%%%%%%%%%%%%%%%%%%%%%%%%%%%%%%%
\section{Theoretical Details for $B \rightarrow M \nu \bar{\nu}~ (M= \pi, K, \rho, K^*)$ Decays}
\label{section3}
%%%%%%%%%%%%%%%%%%%%%%%%%%%%%%%%%%%%%%%%%%
%%%%%%%%%%%%%%%%%%%%%%%%%%%%%%%%%%%%%%%%%%

In the SM, the effective Hamiltonian describing $b\to q \nu\bar{\nu}~ (q=d,s)$
decays is given by
\begin{eqnarray}
H_{\rm eff}(b\to q \nu_{\rm SM} \bar{\nu}_{\rm SM})
 = \frac{G_F \alpha}{2 \pi \sqrt{2}}
          V_{tb} V{_{tq}^\ast} C{_{10} ^\nu}
           \bar{q} \gamma^\mu ( 1- \gamma^5 ) b
           \bar{\nu} \gamma_\mu( 1 - \gamma^5 ) \nu ,
\end{eqnarray}
where $G_F$ is the Fermi constant, $\alpha$ is the fine structure constant
%at the $O(M_Z)$ scale,
and $V_{ij}$ are elements of the CKM matrix. The Wilson coefficient
$C_{10}^\nu$ is dominated by the short-distance dynamics associated
with top quark exchange \cite{Buchalla:2000sk}, and  has the
theoretical uncertainty due to the error of top quark mass, whose
explicit form can be found in literatures
\cite{Aliev:1997se,Buchalla:1995vs,Misiak:1999yg}.
Therefore, aside from the CKM matrix elements, for the $b\to q \nu \bar{\nu}$
decays, there remains only well-controllable theoretical uncertainty
from $C_{10}^\nu$. Although it is not easy to measure experimentally
due to only single meson with the missing energy in the final state,
it can be analyzed through the reconstruction of another $B$ meson
from the $\Upsilon(4S) \to B^+ B^-$ event.

Because the leptophobia means that
the $U(1)^\prime$ charge is zero
for all the ordinary left-handed and right-handed lepton fields within the SM,
the extra leptophobic $Z^\prime$ gives no additional contribution
to $B\to M + E\hspace{-0.25cm}/$ decays if
the missing energy is solely due to the SM neutrinos.
However, $\nu^c$ or $S^c$ in the {\bf 27} representation
of $E_6$, as shown in Table~\ref{table:embedding},
may be interpreted as the right-handed neutrino
and so there can arise additional contributions to the missing energy signal
in $B\to M + E\hspace{-0.25cm}/$ decays.

Here we take $\nu^c$ as a candidate for the right-handed neutrino
and assume that other heavy exotic fermions and bosons be integrated out.
Then, the effective Hamiltonian responsible for $b\to q \nu_R \bar{\nu}_R$
decay is written as
\begin{equation}
H_{\rm eff}(b\to q \nu_R \bar{\nu}_R)
 =  \frac{\pi \alpha}{ \sin^2 2\theta_W M_{Z^\prime}^2}
U_{qb}^{Z^\prime} Q_{\nu_R}^{Z^\prime}
\bar{q}\gamma^\mu (1+\gamma_5) b
\bar{\nu} \gamma_\mu (1+\gamma_5)\nu  ,
\end{equation}
where
\begin{equation}
Q_{\nu_R}^{Z^\prime} = \frac{1}{2}\kappa
\sqrt{ \frac{5 \sin^2 \theta_W}{3}} ,
\end{equation}
and $\kappa$ depends on the details of unification.
Here we take $\kappa=1$ for simplicity.
Then, the total branching ratios (BRs) for
$B\to M + E\hspace{-0.3cm}/\hspace{0.1cm}$
signals are given by
\begin{equation}
{\cal B}(B\to M \nu \bar{\nu}) =
{\cal B}(B\to M \nu_{\rm SM} \bar{\nu}_{\rm SM})
+{\cal B}(B \to M \nu_R \bar{\nu}_R) .
\end{equation}
First we take all the neutrinos massless, then
the differential decay rates for $B \to M \nu \bar{\nu} ~ (M=P,V)$
in the leptophobic $Z^\prime$ model
are given by
\begin{eqnarray}
&&\hspace{-1cm}
\frac{d\Gamma}{d q^2}(B^{\pm} \to P^{\pm}\nu \bar{\nu})
 = \frac{G_F^2 \alpha^2 M_B^3}{2^8 \pi^5} |V_{tb} V{_{tq}^\ast}|^2
       |C_{10}^\nu|^2
\nonumber \\
&&\hspace{-1cm}
 \hspace{3.4cm} \times
\left[
1_{\rm SM}+ \frac{\pi^2 {Q_{\nu_R}^{Z^\prime}}^2 |U_{qb}^{Z^\prime}|^2}
    {\alpha^2 |V_{tb} V{_{tq}^\ast}|^2 |C_{10}^\nu|^2}
    \left( \frac{M_Z^2}{M_{Z^\prime}^2} \right)^2
\right]
\lambda_P^{3/2} |f_+^P(q^2)|^2 ,
\\
&&\hspace{-1cm}
\frac{d\Gamma}{d q^2}(B^{\pm} \to V^{\pm}\nu \bar{\nu})
= \frac{G{_F ^2} \alpha^2 M_B^3}{2^{10} \pi^5}
       |V_{tb} V{_{tq}^*}|^2
        |C{_{10}^\nu}|^2  \lambda_V^{1/2}
\nonumber\\
&&\hspace{-1cm}
 \hspace{3.4cm} \times
\left[
1_{\rm SM}+ \frac{\pi^2 {Q_{\nu_R}^{Z^\prime}}^2 |U_{qb}^{Z^\prime}|^2}
    {\alpha^2 |V_{tb} V{_{tq}^\ast}|^2 |C_{10}^\nu|^2}
    \left( \frac{M_Z^2}{M_{Z^\prime}^2} \right)^2
\right]
\nonumber \\
&&\hspace{-1cm}
 \hspace{3.4cm}
\times \left[ \frac{8 \lambda_V \tilde{q}^2}
         {\left( 1 + \sqrt{r_V} \right)^2} V^V(q^2)^2
       + \frac{1}{r_V}
          \bigg\{ \left(1 + \sqrt{r_V}\right)^2
                \left(\lambda_V +12 r_V \ \tilde{q}^2\right)
                A_1^V(q^2)^2 \right.
\nonumber\\
&&\hspace{-1cm}
 \hspace{3.4cm}
+ \left.  \frac{\lambda_V^2}{\left(1 +\sqrt{r_V}\right)^2} A_2^V(q^2)^2
     -\ 2\ \lambda_V \left( 1 - r_V - \tilde{q}^2 \right)
      {\rm Re} \left(A_1^V(q^2) A_2^V(q^2)^\ast \right)
        \bigg\} \right],
\end{eqnarray}\\
where $\lambda_M=1 +r_M ^2
+\tilde{q}^4 -2r_M -2\tilde{q}^2 -2r_M \tilde{q}^2$
   with $r_M = M_M ^2/M_B ^2, \tilde{q}^2 =q^2/M_B ^2 $
and $P (V)$ in the superscript and subscript
stands for a pseudoscalar (vector) meson, respectively.

Main theoretical uncertainties arise from the hadronic transition form factors
for the $B$ meson decay into the $P(V)$ meson.
However, one can also reduce these uncertainties
by taking ratios of the BRs, related by $SU(3)$ flavor symmetry
or isospin symmetry, such as
$B \to \pi(\rho) \nu \bar{\nu}$ and $B \to \pi(\rho) \ell \nu$
\cite{Aliev:1997se}.
Especially, the corrections to the strict isospin symmetry due to
mass difference between $\rho^\pm$ and $\rho^0$ and form factors
are expected to be very small.

%%%%%%%%%%%%%%%%%%%%%%%%%%%%%%%%%%%%%%%%%%
%%%%%%%%%%%%%%%%%%%%%%%%%%%%%%%%%%%%%%%%%%
\section{Numerical analysis and discussions}
\label{section4}
%%%%%%%%%%%%%%%%%%%%%%%%%%%%%%%%%%%%%%%%%%
%%%%%%%%%%%%%%%%%%%%%%%%%%%%%%%%%%%%%%%%%%

In Table~\ref{table:2}, we show the theoretical estimates of BRs within the SM
and their current experimental bounds at $B$ factories \cite{Abe:2005bq,Aubert:2004ws}.
The errors of the SM estimates in Tab.~\ref{table:2} are
mainly due to the hadronic transition form factors and the
CKM matrix elements:
For the form factors, we use recent results from the LCSR model
with theoretical uncertainties of
about 10 to 13 \% at zero momentum transfer \cite{Ball:2004ye,Ball:2004rg}:
\begin{eqnarray}
 f_+^\pi (0) = 0.258 \pm 0.031 ,&&
~~   f_+^K (0) = 0.331 \pm 0.041 ,
\\
 V^\rho (0) = 0.323 \pm 0.030 ,&&
~~ V^{K^\ast}(0) = 0.411 \pm 0.033 ,
\\
 A_1^\rho (0) = 0.242 \pm 0.023 ,&&
~~ A_1^{K^\ast}(0) = 0.292 \pm 0.028 ,
\\
 A_2^\rho (0) = 0.221 \pm 0.023 ,&&
~~ A_2^{K^\ast}(0) = 0.259 \pm 0.027 .
\end{eqnarray}
The details of the form factors are given in Appendix.
For the CKM matrix elements, we adopted PDG results \cite{Eidelman:2004wy},
\begin{equation}
|V_{tb}| \simeq 1 , ~~ |V_{ts}| = 0.040 \pm 0.003 , ~~
|V_{td}| = (9.4 \pm 4.6)\times 10^{-3} .
\end{equation}

% Table 2
\begin{table}
\caption{Expected BRs in the SM and  experimental bounds (90\% C.L.)
in units of $10^{-6}$.
}
\smallskip
\begin{tabular}{|c|c|c|}
\hline
~~mode~~ & ~~BRs in the SM~~ & ~~Experimental bounds~~ \\
\hline
~~$B\to K\nu\bar{\nu}$~~       & $5.31^{+1.11}_{-1.03}$
                               & $<\phantom{1}36$ \cite{Abe:2005bq} \\
~~$B\to \pi\nu\bar{\nu}$~~     & $0.22^{+0.27}_{-0.17}$
                               & $<100 $ \cite{Aubert:2004ws} \\
~~$B\to K^\ast \nu\bar{\nu}$~~ & $11.15^{+3.05}_{-2.70}$ & - \phantom{-} \\
~~$B\to \rho \nu\bar{\nu}$~~   & $0.49^{+0.61}_{-0.38}$  & - \phantom{-} \\
   \hline
   \end{tabular}
   \label{table:2}
\end{table}

Please note that $b\to d$ transition modes receive much larger theoretical
uncertainties from $|V_{td}|$, as can be seen in Table~\ref{table:2}.
The BRs of $B\to K^{(\ast)} \nu\bar{\nu}$ decays, which are
the $b\to s$ transition process,
are of an order of $10^{-5}$ to $10^{-6}$.
On the other hand, the BRs of $B\to \pi(\rho)\nu\bar{\nu}$
arising from the $b\to d$ transition
are less by about one order.
And as expected from larger degrees of freedom in final states,
BRs of the vector boson production processes are about 2 or 3 times larger
than the corresponding pseudoscalar processes.

Recently the Belle and BaBar  Collaborations
presented the upper limits on BRs of $B\to K\nu \bar{\nu}$
and $B\to \pi \nu \bar{\nu}$ decays \cite{Abe:2005bq,Aubert:2004ws}.
For the $K$ meson production process with the missing energy
the current experimental bound is about 7 times larger than
the SM expectation,
whereas for the $\pi$ meson production the upper bound is much higher
by an order of $10^{3}$ than the theoretical estimate.
Therefore, there is plenty of room for new physics effects in these decays
at present.
The BRs of all the modes are expected to be (more precisely)
measured at $B$ factories soon.

%%%%%%%%%%%%%%%%%%%%%%%%%%%%%%%%%%%%%%%%%%  Fig. 1
%%%%%%%%%%%%%%%%%%%%%%%%%%%%%%%%%%%%%%%%%%
\begin{figure}
\begin{tabular}{cc}
~~~\psfig{file=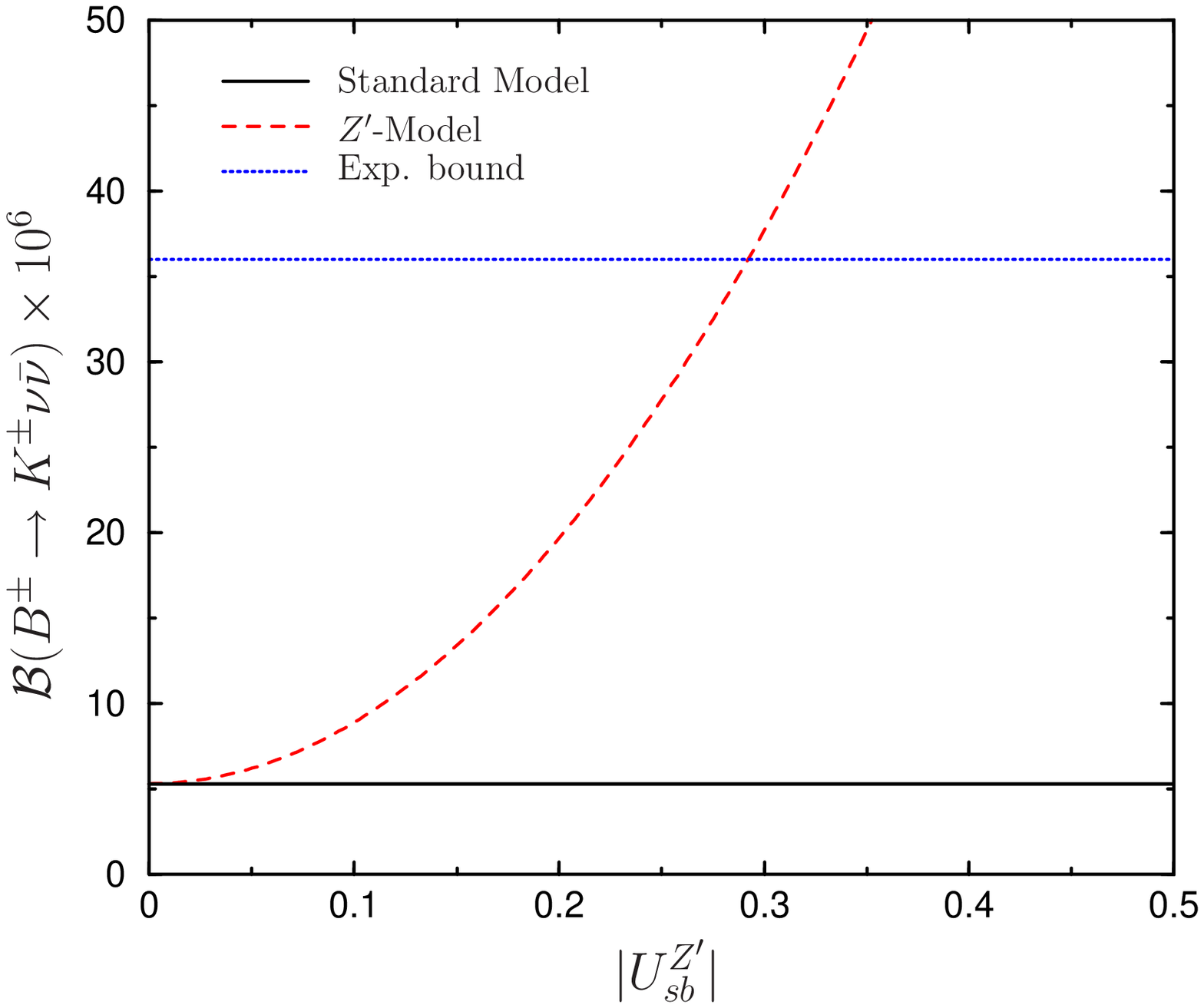,width=7cm}~~~&
~~~\psfig{file=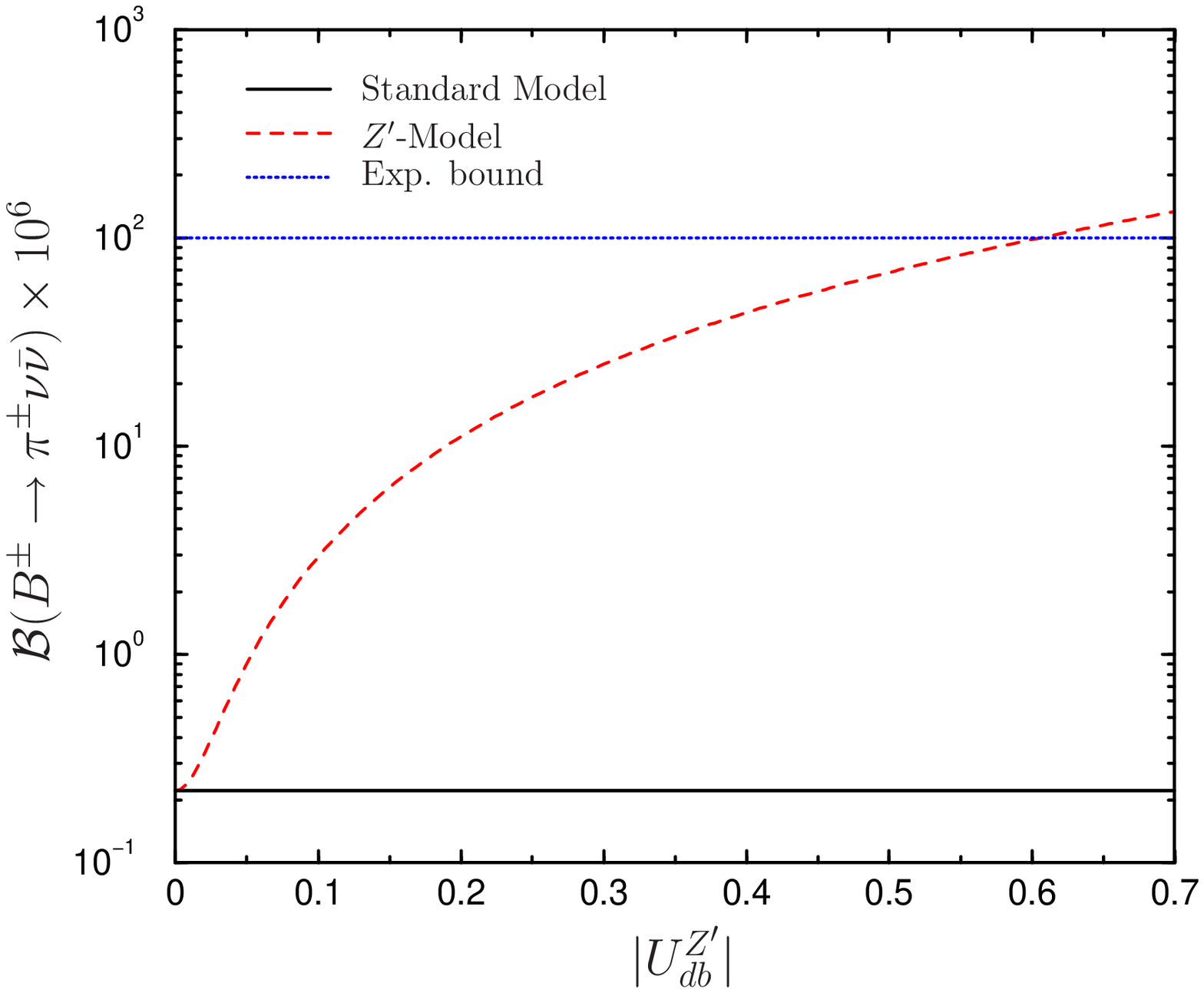,width=7cm}~~~\\[-2.0ex]
\textbf{(a)}&\textbf{(b)}
\end{tabular}
\vspace*{8pt}
\caption{ \label{fig1}
Branching ratios for (a)~$B^\pm \to K^\pm \nu\bar{\nu}$ and
(b)~$B^\pm \to \pi^\pm \nu\bar{\nu}$,
where $\nu$ can be the ordinary SM neutrinos or right-handed neutrinos.
}
\end{figure}
%%%%%%%%%%%%%%%%%%%%%%%%%%%%%%%%%%%%%%%%%%
%%%%%%%%%%%%%%%%%%%%%%%%%%%%%%%%%%%%%%%%%%

%%%%%%%%%%%%%%%%%%%%%%%%%%%%%%%%%%%%%%%%%%  Fig. 2
%%%%%%%%%%%%%%%%%%%%%%%%%%%%%%%%%%%%%%%%%%
\begin{figure}
\begin{tabular}{cc}
~~~\psfig{file=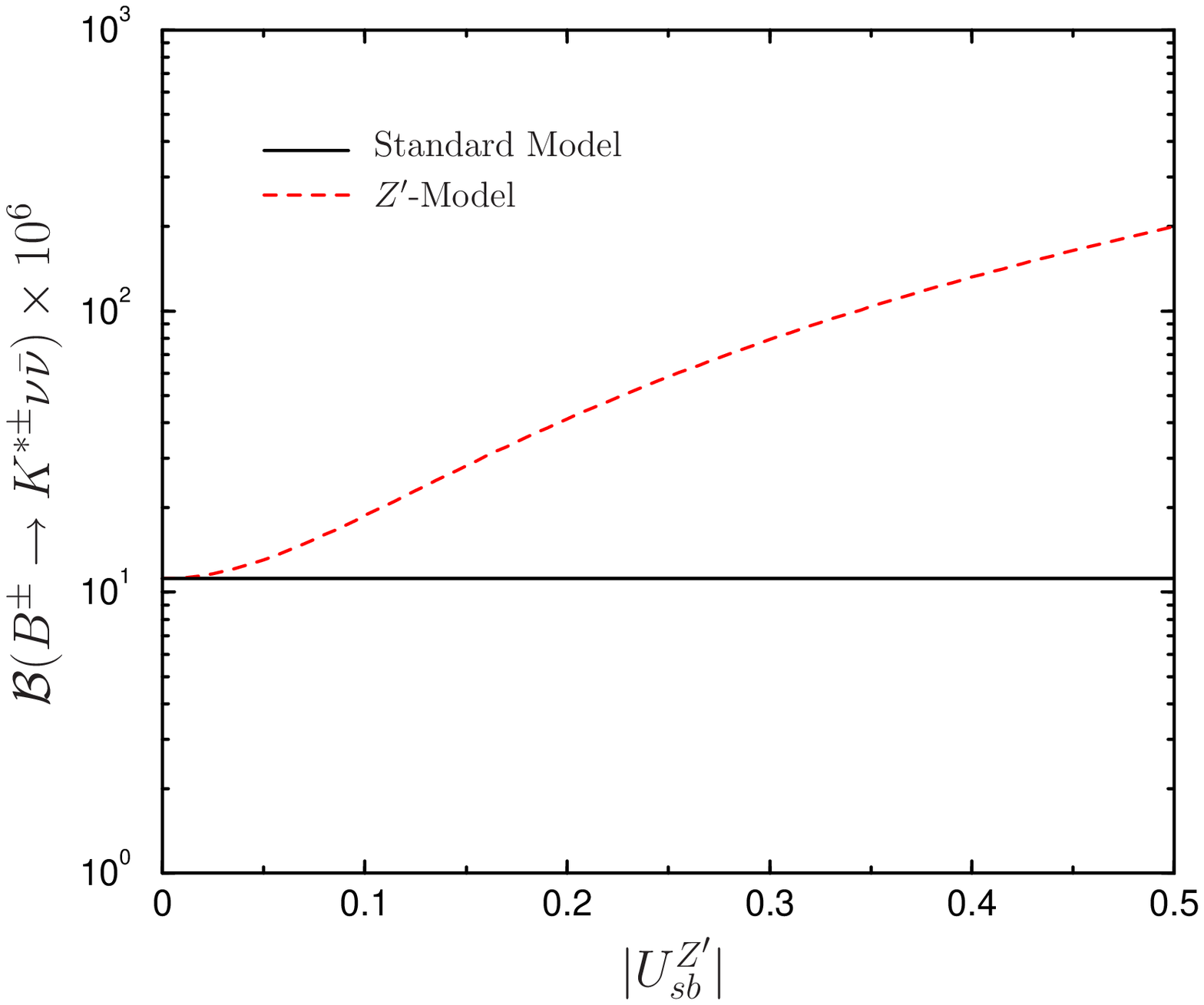,width=7cm}~~~&
~~~\psfig{file=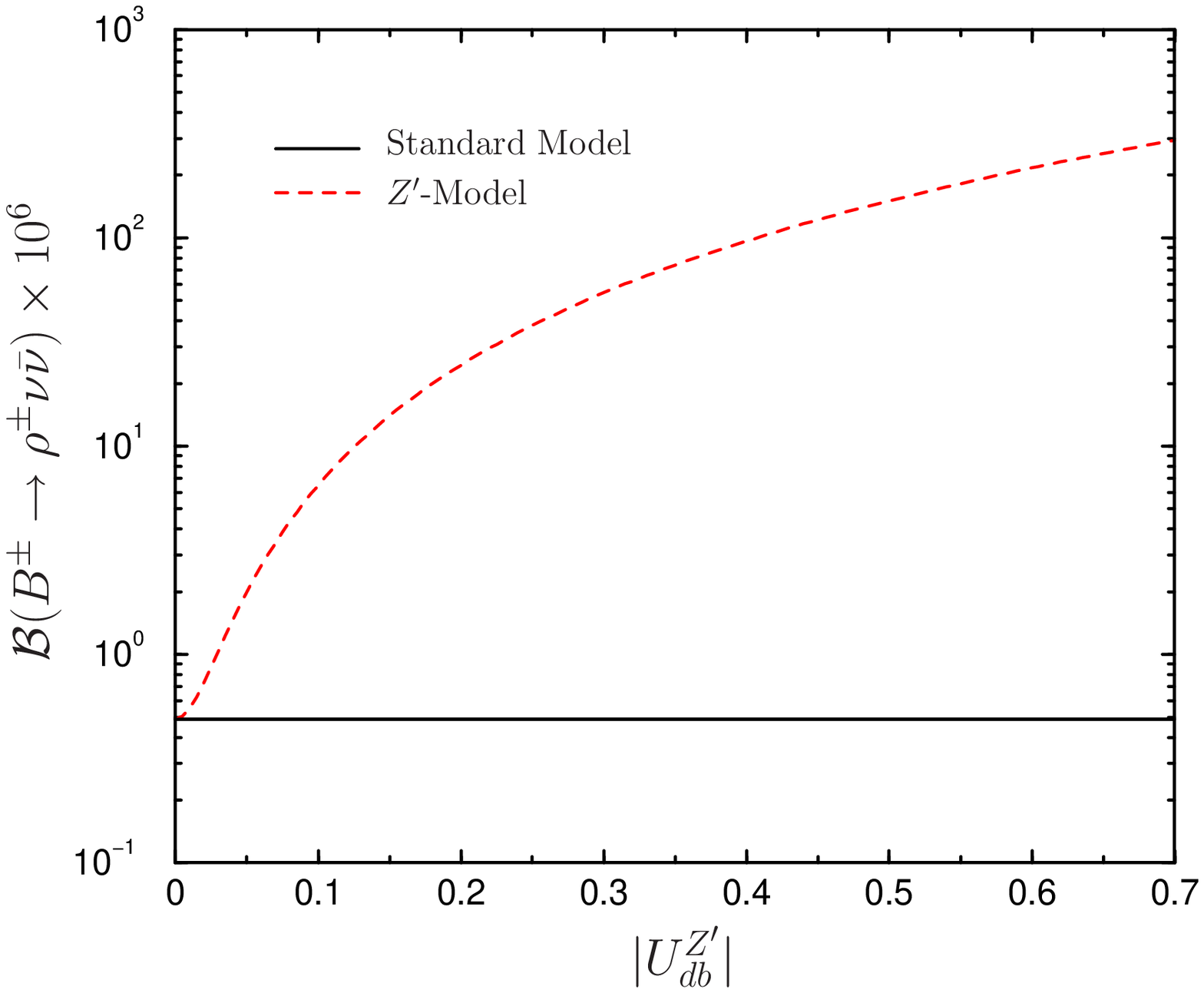,width=7cm}~~~\\[-2.0ex]
\textbf{(a)}&\textbf{(b)}
\end{tabular}
\vspace*{8pt}
\caption{ \label{fig2}
Branching ratios for
(a)~$B^\pm \to K^{\ast\pm} \nu\bar{\nu}$
and (b)~$B^\pm \to \rho^\pm \nu\bar{\nu}$,
where $\nu$ can be the ordinary SM neutrinos or right-handed neutrinos.
}
\end{figure}
%%%%%%%%%%%%%%%%%%%%%%%%%%%%%%%%%%%%%%%%%%
%%%%%%%%%%%%%%%%%%%%%%%%%%%%%%%%%%%%%%%%%%

Now we investigate additional effects from the leptophobic $Z^\prime$ boson.
In Figs.~\ref{fig1} and \ref{fig2}, we present our predictions
for the BRs in the leptophobic $Z^\prime$ model
as a function of the effective coupling $|U_{qb}^{Z^\prime}|$ in the model.
The solid and dotted lines denote the SM predictions and the current experimental
bounds, respectively.
In the leptophobic $Z^\prime$ model, we have two parameters,
the mass of $Z^{\prime}$ boson
and new FCNC coupling, $|U_{qb}^{Z^\prime}|$.
Since the D0 experiment
excludes the mass range $365~{\rm GeV} \leq M_{Z^\prime} \leq 615~{\rm GeV}$
\cite{Abbott:1997dr},
we take $M_{Z^\prime}= 700$ GeV,
which is also consistent with the mass bound of the conventional non-leptophobic $Z^\prime$ model.
Although we choose a specific mass for the $Z^\prime$ boson,
our analysis can be easily translated through the corresponding
changes in the effective coupling $|U_{qb}^{Z^\prime}|$ for different $Z^\prime$ masses.
Considering the current experimental upper bounds,
one can extract the following constraints for the FCNC couplings
from Fig.~\ref{fig1},
\begin{equation}
|U_{sb}^{Z^\prime}| \leq 0.29 , ~~
|U_{db}^{Z^\prime}| \leq 0.61 ,
\end{equation}
for $B\to K \nu\bar{\nu}$ and $B\to \pi \nu \bar{\nu}$ decays, respectively.
Compared with the inclusive $b\to s \nu \bar{\nu}$ decay case \cite{Leroux:2001fx},
we find presently the exclusive mode gives more stringent bounds on the leptophobic
FCNC couplings.
%In addition, the $B\to M \nu \bar{\nu}$ decays are the most promising
%processes for the probe of the FCNC from the leptophobic $Z^\prime$ model.

As already mentioned, although the exclusive $B\to M \nu \bar{\nu}$
decay modes are much easier at the experimental detection than the
inclusive ones, they have inevitable theoretical uncertainties from
hadronic form factors. For example, our expectation value for $B\to
K \nu \bar{\nu}$ in the SM with recent form factors from LCSR model
is about $5.31 \times 10^{-6}$, however, it can be changed to  $\sim
3.8 \times 10^{-6}$ by using different form factors
\cite{Buchalla:2000sk}. In order to reduce the form factor
uncertainties, we take ratios for ${\cal B}(B\to M \nu \bar{\nu})$
to ${\cal B}(B\to M e \nu)$ for $M=\pi,\rho$ mesons. In the limit of
isospin symmetry, and requiring $M_\pi^\pm = M_\pi^0, M_\rho^\pm =
M_\rho^0$, we obtain
\begin{eqnarray}
&&\frac{{\cal B}(B^{\pm} \rightarrow \pi^{\pm} \nu \bar{\nu})}
     {{\cal B}(B^{\pm} \rightarrow \pi^0 e^\pm \nu)}
=  \frac{3 \alpha^2}{2 \pi^2}
    \left| \frac{V_{td}}{V_{ub}} \right|^2
    |C_{10}^\nu|^2
+   \frac{3}{2}
    \frac{{Q_{\nu_R}^{Z^\prime}}^2 | U_{db}^{Z^\prime}|^2}
    {|V_{ub}|^2}
    \left( \frac{M_Z^2}{M_{Z^\prime}^2} \right)^2
\nonumber \\
&&
= 2.5\times 10^{-3} \left( \frac{0.40}{|V_{ub}/V_{td}|}\right)^2
 + 7.6\times 10^{-3} \left( \frac{3.7\times 10^{-3}}{|V_{ub}|}\right)^2
 \left( \frac{700~\rm{GeV} }{M_{Z^\prime}}\right)^4
 \left( \frac{|U_{db}^{Z^\prime}|}{0.05}\right)^2 ,
\\
&&\frac{{\cal B}(B^{\pm} \rightarrow \rho^{\pm} \nu \bar{\nu})}
     {{\cal B}(B^{\pm} \rightarrow \rho^0 e^\pm \nu)}
=  \frac{3 \alpha^2}{2 \pi^2}
    \left| \frac{V_{td}}{V_{ub}} \right|^2
    |C_{10}^\nu|^2
+  \frac{3}{2}
   \frac{{Q_{\nu_R}^{Z^\prime}}^2 | U_{db}^{Z^\prime}|^2}
    {|V_{ub}|^2}
    \left( \frac{M_Z^2}{M_{Z^\prime}^2} \right)^2
\nonumber \\
&&
= 2.5\times 10^{-3} \left( \frac{0.40}{|V_{ub}/V_{td}|}\right)^2
 + 7.6\times 10^{-3} \left( \frac{3.7\times 10^{-3}}{|V_{ub}|}\right)^2
 \left( \frac{700~\rm{GeV} }{M_{Z^\prime}}\right)^4
 \left( \frac{|U_{db}^{Z^\prime}|}{0.05}\right)^2 ,
\end{eqnarray}
where the form factor dependencies are all cancelled out.
For $K^{(\ast)}$ meson productions, however,
since there does not exist such a corresponding semi-leptonic decay mode,
one cannot adopt the same analysis given above.
Instead, considering $M_\rho \approx M_{K^\ast}$ and the flavor $SU(3)$ symmetry,
we obtain the following relation:
\begin{eqnarray}
&& \hspace{-0.5cm}
\frac{{\cal B}(B^{\pm} \rightarrow K^{\ast \pm} \nu \bar{\nu})}
     {{\cal B}(B^{\pm} \rightarrow \rho^0 e^\pm \nu)}
   \simeq
    \frac{3 \alpha^2}{2 \pi^2}
    \left| \frac{V_{ts}}{V_{ub}} \right|^2
    |C_{10}^\nu|^2
+ \frac{3}{2}
 \frac{{Q_{\nu_R}^{Z^\prime}}^2 | U_{sb}^{Z^\prime}|^2}
    {|V_{ub}|^2}
    \left( \frac{M_Z^2}{M_{Z^\prime}^2} \right)^2
\nonumber \\
&&
= 4.5\times 10^{-2} \left( \frac{9.3\times 10^{-2}}{|V_{ub}/V_{ts}|}\right)^2
 + 1.2\times 10^{-1} \left( \frac{3.7\times 10^{-3}}{|V_{ub}|}\right)^2
 \left( \frac{700 ~\rm{GeV}}{M_{Z^\prime}}\right)^4
 \left( \frac{|U_{sb}^{Z^\prime}|}{0.2}\right)^2 .
\end{eqnarray}
On the contrary to $B^\pm \to K^\ast \nu\bar{\nu}$
and $B^\pm \to \rho^0 e^\pm \nu$,
we cannot use the $SU(3)$ flavor symmetry between
$B^\pm \to K^\pm  \nu\bar{\nu}$ and
$B^\pm \to \pi^0 e^\pm \nu$ transitions because of the possibly large $SU(3)$
breaking due to  large mass difference
between $\pi$ and $K$ mesons.

%%%%%%%%%%%%%%%%%%%%%%%%%%%%%%%%%%%%%%%%%%  Fig. 3
%%%%%%%%%%%%%%%%%%%%%%%%%%%%%%%%%%%%%%%%%%
\begin{figure}
\begin{tabular}{cc}
\psfig{file=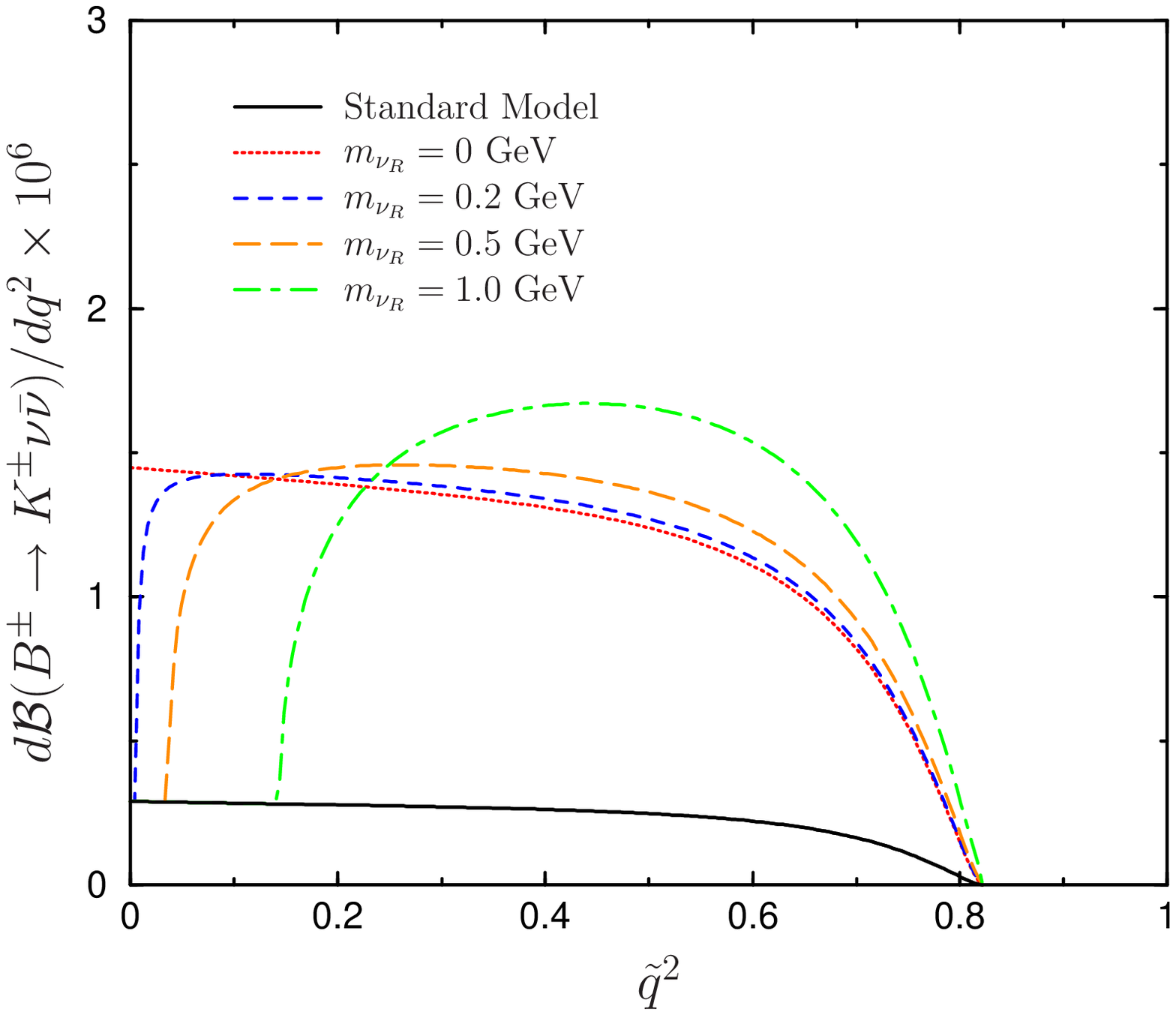,width=7cm}&~~
\psfig{file=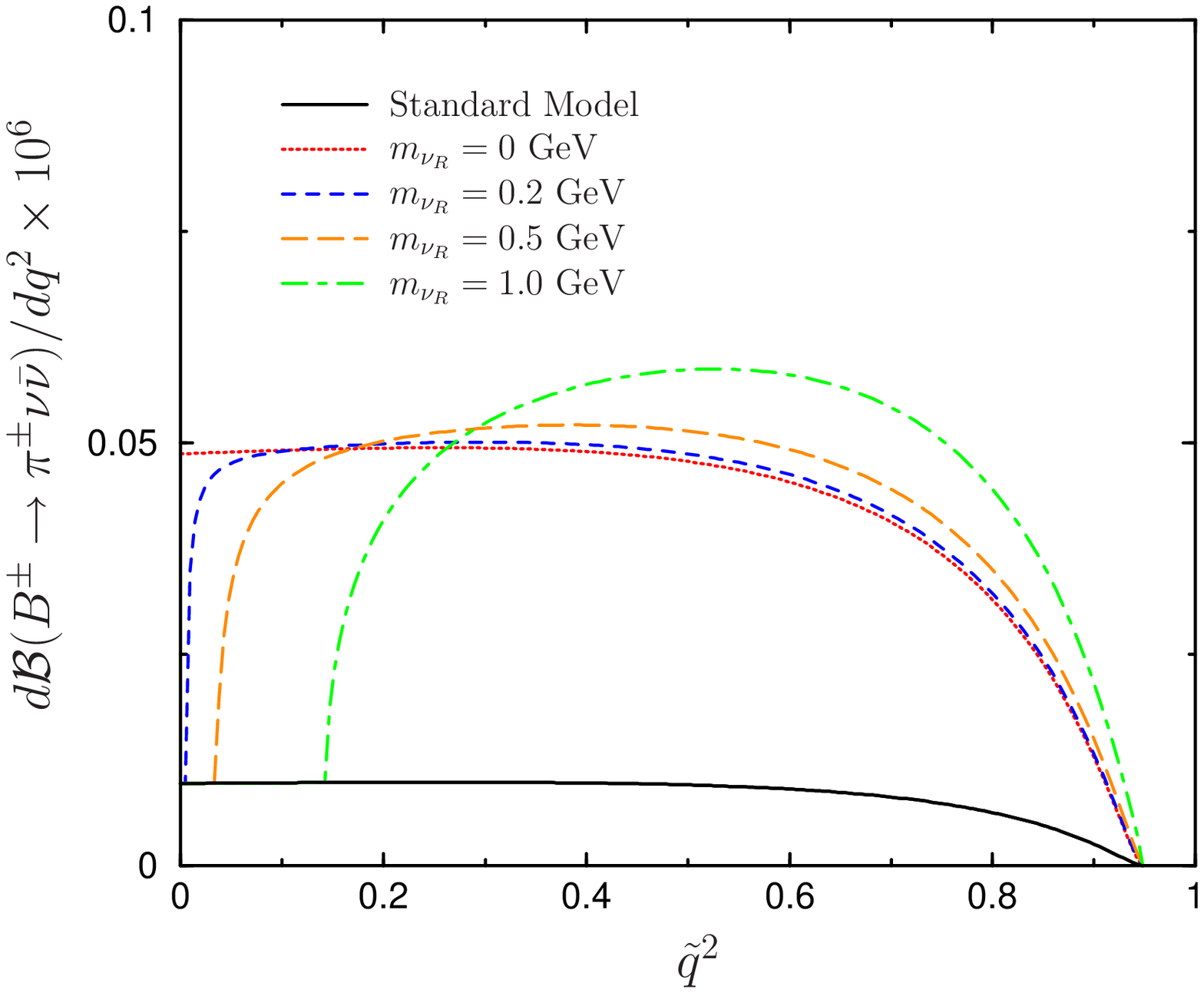,width=7cm}\\[-2.0ex]
\textbf{(a)}&~~\textbf{(b)}\\
\psfig{file=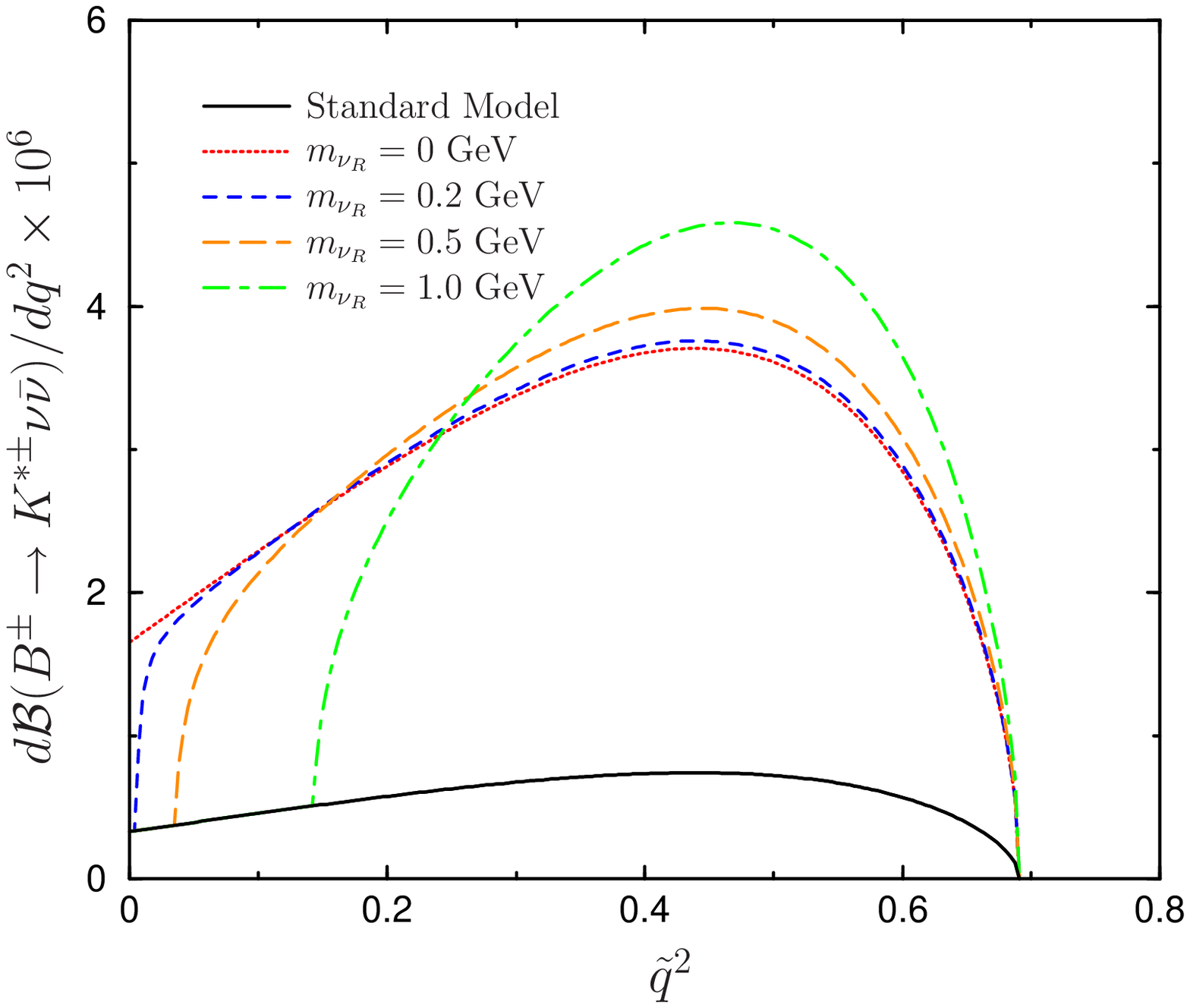,width=7cm}&~~
\psfig{file=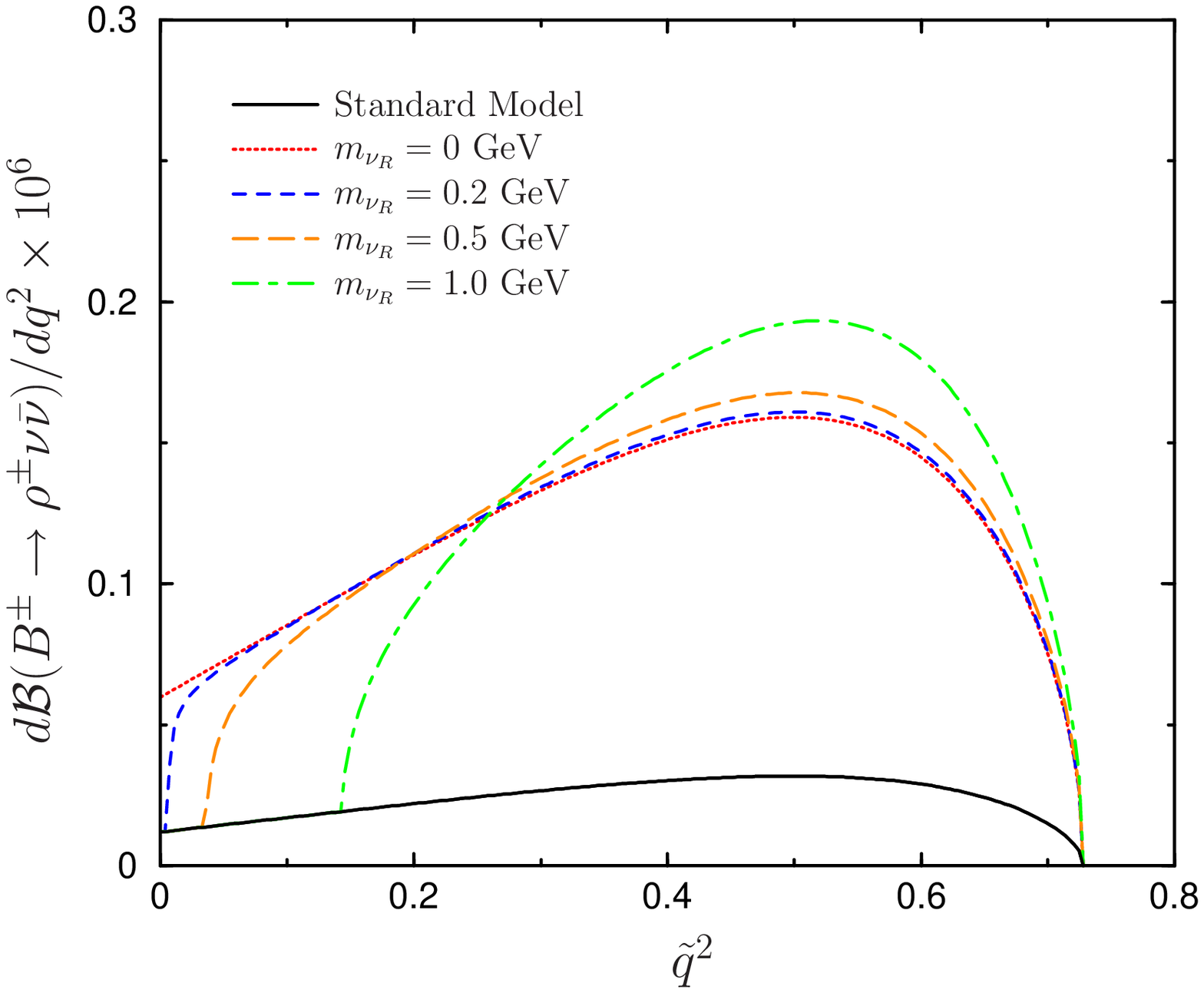,width=7cm}\\[-2.0ex]
\textbf{(c)}&~~\textbf{(d)}
\end{tabular}
\vspace*{8pt}
\caption{ \label{fig3}
Differential BRs as a function of
the normalized momentum transfer square, $\tilde{q}^2=q^2/M_B^2$, in units of
$10^{-6}$, for (a)~$B^\pm \to K^\pm \nu\bar{\nu}$,
(b)~$B^\pm \to \pi^\pm \nu\bar{\nu}$,
(c)~$B^\pm \to K^{\ast\pm} \nu\bar{\nu}$,
and (d)~$B^\pm \to \rho^\pm \nu\bar{\nu}$.
Here the decay rates in the leptophobic $Z^\prime$ model
are normalized to be five times larger than those in the SM.
}
\end{figure}
%%%%%%%%%%%%%%%%%%%%%%%%%%%%%%%%%%%%%%%%%%
%%%%%%%%%%%%%%%%%%%%%%%%%%%%%%%%%%%%%%%%%%

Up to now we have assumed  all the neutrinos are massless.
Aside from the SM neutrinos of which masses are indeed very small,
the extra right-handed neutrinos in the leptophobic $Z^\prime$ model
can have sizable masses.
Effects of the massive neutrinos will be an interesting subject by themselves.
Here we briefly introduce the mass effects.
In Fig.~\ref{fig3}, we present
the differential BRs as a function of the normalized momentum transfer-squared
for $m_{\nu_R}=0,0.2,0.5,1~{\rm GeV}$.
%We fixed the parameters as:
%$$
%M_{Z^\prime}= 700~~{\rm GeV}, ~~~~~~
%|U_{sb}^{Z^\prime}| = 0.29, ~~~~~~
%|U_{db}^{Z^\prime}| = 0.61 .
%$$
We normalized the integrated BRs such that the BRs in the leptophobic $Z^\prime$ model
are five times larger than those in the SM.
It is apparent that the differential BRs show the specific features
according to the corresponding mass of the right-handed neutrino.
Especially the sharp rise near the threshold point  allows
to increase the accuracy of the mass measurement of
the lightest right-handed neutrino.
Unfortunately, if its mass is lower than a few hundred MeV,
it is very hard to find the difference from the massless neutrino case.

In general, the additional right-handed neutrinos
can have Dirac mass terms coupled with the ordinary
SM neutrinos as well as Majorana masses \cite{Langacker:1988ur}.
Since there is no limit for both Dirac and Majorana masses
in principle, they are free parameters.
One interesting scenario is a so-called ``{$\nu$}MSM'' model \cite{Asaka:2005pn}, which
allows three right-handed (light) neutrinos in addition to the SM ones.
Assuming that Majorana masses of right-handed neutrinos are
the order of the electroweak scale or below and
Yukawa couplings are very small,
this model can be accommodated with the present neutrino data.
Especially, if masses of the right-handed neutrinos
are in range of $2\lesssim M_{\nu_R} \lesssim 5~{\rm keV}$,
they can be good candidates for the warm dark matters,
and they can explain the neutrino oscillation via the the see-saw mechanism
and baryon asymmetry of the universe as well \cite{Asaka:2005pn}.

To conclude,
the FCNC processes with the missing energy signal in $B$ decays
have been studied in the leptophobic $Z^\prime$ model.
In general the direct probe of the leptophobic $Z^\prime$ model
may be very difficult at ILC as well as LHC because of large hadronic  backgrounds.
In this regard the exclusive FCNC processes $B\to M \nu \bar{\nu}$
are very adequate to measure  new physics
in the leptophobic $Z^\prime$ model with the right-handed neutrinos.
This model could be quite important in the context of possibly large new physics
scenario in the EW penguin sector.
We also showed that  ratios of BRs can reduce the large hadronic
uncertainty from form factors.
The differential BRs are very useful if the right-handed neutrinos
have the sub-GeV masses.
We also noted that the right-handed neutrinos could be accommodated with
the ``$\nu$MSM'' scenario.

%%%%%%%%%%%%%%%%%%%%%%%%%%%%%%%%%%%%%%%%%%%%%%%%%%%%%%
%%%%%%%%%%%%%%%%%%%%%%%%%%%%%%%%%%%%%%%%%%%%%%%%%%%%%%
\vspace{1cm} \centerline{\bf ACKNOWLEDGMENTS}
\noindent We thank
Y.G. Kim and U. Haisch for careful reading and useful comments. The
work of J.H.J. was supported  by
the Korea Research Foundation Grant funded by the Korean
Government (MOEHRD) No. R02-2003-000-10050-0. The work of C.S.K. was
supported in part by  CHEP-SRC Program and in part by the Korea
Research Foundation Grant funded by the Korean Government (MOEHRD)
No. KRF-2005-070-C00030. The work of
J.L. was supported in part by Brain Korea 21 Program and in part by
Grant No. F01-2004-000-10292-0 of KOSEF-NSFC International
Collaborative Research Grant. The work of C.Y. was supported by the
Korea Research Foundation Grant funded by Korea Government (MOEHRD,
Basic Research Promotion FUND) (KRF-2005-075-C00008).
\\

%%%%%%%%%%%%%%%%%%%%%%%%%%%%%%%%%%%%%%%%%%%%%%%%%%%%%%
%%%%%%%%%%%%%%%%%%%%%%%%%%%%%%%%%%%%%%%%%%%%%%%%%%%%%%
\section*{APPENDIX : HADRONIC FORM FACTORS}
%%%%%%%%%%%%%%%%%%%%%%%%%%%%%%%%%%%%%%%%%%
%%%%%%%%%%%%%%%%%%%%%%%%%%%%%%%%%%%%%%%%%%

The hadronic matrix elements for $B \rightarrow P \nu \bar{\nu}$ ($P = \pi, K$)
decays can be parameterized in terms of the form factors $f_+^P(q^2)$ and
$f_-^P(q^2)$:
\begin{eqnarray}
\langle P(p_2)| \bar{q} \gamma^\mu (1 -\gamma^5 ) b | B(p_1) \rangle
&=& p^\mu f_+^P(q^2) + q^\mu f_-^P(q^2) ,
\end{eqnarray}
where $p=p_1 + p_2$ and  $q=p_1 -p_2$.
For $B \rightarrow V \nu \bar{\nu}$ ($V = \rho, K^*$)decays,
the hadronic matrix element can be written in terms of four form factors as
\begin{eqnarray}
&&\langle V(p_2, \varepsilon)| \bar{q} \gamma^\mu (1 -\gamma^5 )b|B(p_1)\rangle
\nonumber\\
&& \hspace{1cm}
=  - \varepsilon_{\mu \nu \rho \sigma}
       \varepsilon^{* \nu}
       p{_2 ^\rho}
       q^\sigma
       \frac{2V^V(q^2)}{M_B + M_V}
   - i \bigg\{\varepsilon{_\mu ^*}
              \big(M_B +M_V\big)
              A_1^V(q^2)
\nonumber\\
&& \hspace{1.4cm}
             -\big(\varepsilon^* \hspace{-0.1cm}\cdot q\big)
              \big(p_1 +p_2\big)_\mu
              \frac{A_2^V(q^2)}{M_B + M_V}
             -q_\mu
              \big(\varepsilon^* \hspace{-0.1cm}\cdot q\big)
              \frac{2 M_V}{q^2}
              \big(A_3^V(q^2) - A_0^V(q^2)\big)
       \bigg\} ,
\end{eqnarray}
with $A_3^V(0)=A_0^V(0)$ and
     $A_3^V(q^2) = \displaystyle \frac{M_B +M_V}{2M_V}A_1^V(q^2)
              -\frac{M_B -M_V}{2M_V}A_2^V(q^2)$.

% Table 3
\begin{table}
\caption{The parameters for form factors in LCSR model
\cite{Ball:2004ye,Ball:2004rg}
where $m_b=4.8$ GeV is fixed.}
\label{table3}
\smallskip
\begin{tabular}{|c||c|c|c|c|}
\hline
~~parameter~~  & $r_1$  & $r_2$  & $m_{\rm fit}^2$ & $m_R^2$ \\
\hline
 $f_+^K$   &~~ $0.1616$~~ & ~~$0.1730$~~ &  $-$   &~~$5.41^2$~~ \\
 $f_0^K$   &~~ $-$~~      & ~~$0.3302$~~ &  $37.46$   &~~$-$~~ \\
 $f_+^\pi$   &~~ $0.744$~~ & ~~$-0.486$~~ &  $40.73$   &~~$5.32^2$~~ \\
 $f_0^\pi$   &~~ $-$~~      & ~~$0.258$~~ &  $33.81$   &~~$-$~~ \\
\hline
 $V^{K^\ast}$  & $0.923$ & $-0.511$ & ~~ $49.40$~~ & $5.32^2$\\
 $A_0^{K^\ast}$  & $1.364$ & $-0.990$ & ~~ $36.78$~~ & $5.28^2$\\
 $A_1^{K^\ast}$  & $-$   &  $0.290$ &  $40.38$ &$-$\\
 $A_2^{K^\ast}$  &$-0.084$ &  $0.342$ &  $52.00$ &$-$\\
\hline
 $V^\rho$   & $1.045$ & $-0.721$ &  $38.34$ & $5.32^2$\\
 $A_0^\rho$   & $1.527$ & $-1.220$ &  $33.36$ & $5.28^2$\\
 $A_1^\rho$   &  $-$  & $0.240$  &  $37.51$ &$-$\\
 $A_2^\rho$   & $0.009$ & $0.212$  &  $40.82$ &$-$\\
\hline
\end{tabular}
\end{table}
%%%%%%%%%%%%%%%%%%%%%%%%%%%%%%%%%%%%%%%%%%
%%%%%%%%%%%%%%%%%%%%%%%%%%%%%%%%%%%%%%%%%%

For the numerical analysis, we follow the theoretical estimates from LCSR
model \cite{Ball:2004ye,Ball:2004rg}.
The transition form factors are parameterized by three independent
fit parameters and one resonance mass.
Form factors used in the numerical calculation are given by
\begin{eqnarray}
&&    f_+^\pi (q^2) =
\frac{r_1}{1 -q^2/m_R ^2} + \frac{r_2}{1 -q^2/m{_{\text{fit}} ^2}} ,
\\
&&  f_+^K (q^2) =
\frac{r_1}{1 -q^2/m_R ^2} + \frac{r_2}{(1 -q^2/m_R ^2)^2} ,
\\
&&    f_0^P (q^2) =
\frac{r_2}{1 -q^2/m{_{\text{fit}} ^2}} ,
\\
&& F^V(q^2) = \frac{r_1}{1 -q^2 /m_R ^2}
            +\frac{r_2}{1 -q^2 /m_{\rm fit}^2} ~~~({\rm for~} F=V~{\rm or}~
  A_0) ,
\\
&&  A_1^V(q^2) =\frac{r_2}{1 -q^2/m_{\rm fit}^2} ,
\\
&&  A_2^V(q^2) = \frac{r_1}{1 -q^2/m_{\rm fit}^2}
  + \frac{r_2}{(1 -q^2/m_{\rm fit}^2)^2} ,
\end{eqnarray}
where values of parameters are shown in Table~\ref{table3}
and $f_-^P = (f_0^P - f_+^P) \frac{M_B^2-M_P^2}{q^2}$.

\vspace{5cm}
%%%%%%%%%%%%%%%%%%%%%%%%%%%%%%%%%%%%%%%%%%%%%%%%%%%%%%
%%%%%%%%%%%%%%%%%%%%%%%%%%%%%%%%%%%%%%%%%%%%%%%%%%%%%%

%\newpage


\begin{thebibliography}{99}
%\cite{Beneke:1999br}
\bibitem{Beneke:1999br}
  M.~Beneke, G.~Buchalla, M.~Neubert and C.~T.~Sachrajda,
  %``{QCD} factorization for B $\to$ pi pi decays: Strong phases and CP
  %violation in the heavy quark limit,''
  Phys.\ Rev.\ Lett.\  {\bf 83}, 1914 (1999)
  [arXiv:hep-ph/9905312];
   %%CITATION = HEP-PH 9905312;%%
  %``QCD factorization for exclusive, non-leptonic B meson decays: General
  %arguments and the case of heavy-light final states,''
  Nucl.\ Phys.\ B {\bf 591}, 313 (2000)
  [arXiv:hep-ph/0006124];
  %%CITATION = HEP-PH 0006124;%%
   %``QCD factorization in B $\to$ pi K, pi pi decays and extraction of
  %Wolfenstein parameters,''
  Nucl.\ Phys.\ B {\bf 606}, 245 (2001)
  [arXiv:hep-ph/0104110].
  %%CITATION = HEP-PH 0104110;%%

\bibitem{Keum:2000wi}
  Y.~Y.~Keum, H.~N.~Li and A.~I.~Sanda,
  %``Penguin enhancement and B $\to$ K pi decays in perturbative QCD,''
  Phys.\ Rev.\ D {\bf 63}, 054008 (2001)
  [arXiv:hep-ph/0004173];
  %%CITATION = HEP-PH 0004173;%%
  %``Fat penguins and imaginary penguins in perturbative QCD,''
  Phys.\ Lett.\ B {\bf 504}, 6 (2001)
  [arXiv:hep-ph/0004004].
  %%CITATION = HEP-PH 0004004;%%

%\cite{Bauer:2005kd}
\bibitem{Bauer:2005kd}
  C.~W.~Bauer, I.~Z.~Rothstein and I.~W.~Stewart,
  %``SCET analysis of B $\to$ K pi, B $\to$ K anti-K, and B $\to$ pi pi
  %decays,''
  arXiv:hep-ph/0510241.
  %%CITATION = HEP-PH 0510241;%%

%\cite{Li:2005kt}
\bibitem{Li:2005kt}
  H.~n.~Li, S.~Mishima and A.~I.~Sanda,
  %``Resolution to the B $\to$ pi K puzzle,''
  Phys.\ Rev.\ D {\bf 72}, 114005 (2005)
  [arXiv:hep-ph/0508041].
  %%CITATION = HEP-PH 0508041;%%

%\cite{Mishima:2004um}
\bibitem{Mishima:2004um}
  S.~Mishima and T.~Yoshikawa,
  %``Large electroweak penguin contribution in B $\to$ K pi and pi pi decay
  %modes,''
  Phys.\ Rev.\ D {\bf 70}, 094024 (2004)
  [arXiv:hep-ph/0408090];
  %%CITATION = HEP-PH 0408090;%%
  Y.~L.~Wu and Y.~F.~Zhou,
  %``Implications of charmless B decays with large direct CP violation,''
  {\it ibid.} {\bf 71}, 021701 (2005)
  [arXiv:hep-ph/0409221];
  %%CITATION = HEP-PH 0409221;%%
  %``Charmless decays B $\to$ pi pi, pi K and K K in broken SU(3) symmetry,''
  {\it ibid.} {\bf 72}, 034037 (2005)
  [arXiv:hep-ph/0503077];
  %%CITATION = HEP-PH 0503077;%%
  A.~J.~Buras, R.~Fleischer, S.~Recksiegel and F.~Schwab,
  %``The B $\to$ pi pi, pi K puzzles in the light of new data: Implications for
  %the standard model, new physics and rare decays,''
  Acta Phys.\ Polon.\ B {\bf 36} (2005) 2015
  [arXiv:hep-ph/0410407];
  %%CITATION = HEP-PH 0410407;%%
  %``Electroweak Penguin Hunting Through B $\to$ pi pi, pi K and Rare K and B
  %Decays,''
  arXiv:hep-ph/0512059;
  %%CITATION = HEP-PH 0512059;%%
  Y.~Y.~Charng and H.~n.~Li,
  %``Weak phases from the B $\to$ pi pi, K pi decays,''
  Phys.\ Rev.\ D {\bf 71}, 014036 (2005)
  [arXiv:hep-ph/0410005];
  %%CITATION = HEP-PH 0410005;%%
  X.~G.~He and B.~H.~J.~McKellar,
  %``Hadron decay amplitudes from B $\to$ K pi and B $\to$ pi pi decays,''
  arXiv:hep-ph/0410098;
  %%CITATION = HEP-PH 0410098;%%
  S.~Baek, P.~Hamel, D.~London, A.~Datta and D.~A.~Suprun,
  %``The B $\to$ pi K puzzle and new physics,''
  Phys.\ Rev.\ D {\bf 71}, 057502 (2005)
  [arXiv:hep-ph/0412086];
  %%CITATION = HEP-PH 0412086;%%
  C.~S.~Kim, S.~Oh and C.~Yu,
  %``A critical study of the B $\to$ K pi puzzle,''
  {\it ibid.} {\bf 72}, 074005 (2005)
  [arXiv:hep-ph/0505060];
  %%CITATION = HEP-PH 0505060;%%
  X.~q.~J.~Li and Y.~d.~J.~Yang,
  %``Revisiting B $\to$ pi pi, pi K decays in QCD factorization approach,''
  Phys.\ Rev.\ D {\bf 72}, 074007 (2005)
  [arXiv:hep-ph/0508079];
  %%CITATION = HEP-PH 0508079;%%
  D.~Chang, C.~S.~Chen, H.~Hatanaka and C.~S.~Kim,
  %``Generalized study with isospin-phased topological approach on the B $\to$ K
  %pi puzzle,''
  arXiv:hep-ph/0510328.
  %%CITATION = HEP-PH 0510328;%%

%\cite{Arnowitt:2005qz}
\bibitem{Arnowitt:2005qz}
  R.~Arnowitt, B.~Dutta, B.~Hu and S.~Oh,
  %``The B $\to$ K pi puzzle and supersymmetric models,''
  Phys.\ Lett.\ B {\bf 633}, 748 (2006)
  [arXiv:hep-ph/0509233].
  %%CITATION = HEP-PH 0509233;%%
  W.~S.~Hou, M.~Nagashima and A.~Soddu,
  %``Difference in B+ and B0 direct CP asymmetry as effect of a fourth
  %generation,''
  Phys.\ Rev.\ Lett.\  {\bf 95}, 141601 (2005)
  [arXiv:hep-ph/0503072];
  %``Enhanced K(L) $\to$ pi0 nu anti-nu from direct CP violation in B $\to$ K pi
  %with four generations,''
  Phys.\ Rev.\ D {\bf 72}, 115007 (2005)
  [arXiv:hep-ph/0508237];
  %%CITATION = HEP-PH 0508237;%%
  Y.~D.~Yang, R.~Wang and G.~R.~Lu,
  %``The puzzles in B $\to$ pi pi and pi K decays: Possible implications for
  %R-parity violating supersymmetry,''
  {\it ibid.} {\bf 73}, 015003 (2006)
  [arXiv:hep-ph/0509273].

%\cite{Barger:2003hg}
\bibitem{Barger:2003hg}
  V.~Barger, C.~W.~Chiang, P.~Langacker and H.~S.~Lee,
  %``Z' mediated flavor changing neutral currents in B meson decays,''
  Phys.\ Lett.\ B {\bf 580}, 186 (2004)
  [arXiv:hep-ph/0310073];
  %%CITATION = HEP-PH 0310073;%%
  %``Solution to the B $\to$ pi K puzzle in a flavor-changing Z' model,''
  {\it ibid}. {\bf 598}, 218 (2004)
  [arXiv:hep-ph/0406126];
  %%CITATION = HEP-PH 0406126;%%
  C.~H.~Chen and H.~Hatanaka,
  %``Nonuniversal $Z^{\prime}$ couplings in $B$ decays,''
  arXiv:hep-ph/0602140.
  %%CITATION = HEP-PH 0602140;%%

%\cite{Grossman:1995gt}
\bibitem{Grossman:1995gt}
  Y.~Grossman, Z.~Ligeti and E.~Nardi,
  %``First limit on inclusive B $\to$ X/s neutrino antineutrino decay and
  %constraints on new physics,''
  Nucl.\ Phys.\ B {\bf 465}, 369 (1996)
  [Erratum-ibid.\ B {\bf 480}, 753 (1996)]
  [arXiv:hep-ph/9510378].
  %%CITATION = HEP-PH 9510378;%%

%\cite{Buchalla:2000sk}
\bibitem{Buchalla:2000sk}
  G.~Buchalla, G.~Hiller and G.~Isidori,
  %``Phenomenology of non-standard Z couplings in exclusive semileptonic  b
  %$\to$ s transitions,''
  Phys.\ Rev.\ D {\bf 63}, 014015 (2001)
  [arXiv:hep-ph/0006136].
  %%CITATION = HEP-PH 0006136;%%

%\cite{Aliev:1997se}
\bibitem{Aliev:1997se}
  T.~M.~Aliev and C.~S.~Kim,
  %``Measuring $|$V(td)/V(ub)$|$ through B $\to$ M nu anti-nu (M = pi,K,rho,K*)
  %decays,''
  Phys.\ Rev.\ D {\bf 58}, 013003 (1998)
  [arXiv:hep-ph/9710428];
  %%CITATION = HEP-PH 9710428;%%
  C.~S.~Kim, Y.~G.~Kim and T.~Morozumi,
  %``New physics effects in B $\to$ K(*) nu nu decays,''
  {\it ibid.} {\bf 60}, 094007 (1999)
  [arXiv:hep-ph/9905528].
  %%CITATION = HEP-PH 9905528;%%

%\cite{Aliev:2001in}
\bibitem{Aliev:2001in}
  T.~M.~Aliev, A.~Ozpineci and M.~Savci,
  %``Rare B $\to$ K* nu anti-nu decay beyond standard model,''
  Phys.\ Lett.\ B {\bf 506}, 77 (2001)
  [arXiv:hep-ph/0101066].
  %%CITATION = HEP-PH 0101066;%%

%\cite{Bobeth:2001jm}
\bibitem{Bobeth:2001jm}
  C.~Bobeth, A.~J.~Buras, F.~Kruger and J.~Urban,
  %``QCD corrections to anti-B $\to$ X/d,s nu anti-nu, anti-B/d,s $\to$ l+ l-,
  %K $\to$ pi nu anti-nu and K(L) $\to$ mu+ mu- in the MSSM,''
  Nucl.\ Phys.\ B {\bf 630}, 87 (2002)
  [arXiv:hep-ph/0112305].
  %%CITATION = HEP-PH 0112305;%%

%\cite{Hewett:1988xc}
\bibitem{Hewett:1988xc}
  J.~L.~Hewett and T.~G.~Rizzo,
  %``Low-Energy Phenomenology Of Superstring Inspired E(6) Models,''
  Phys.\ Rept.\  {\bf 183}, 193 (1989).
  %%CITATION = PRPLC,183,193;%%

%\cite{Leike:1998wr}
\bibitem{Leike:1998wr}
  A.~Leike,
  %``The phenomenology of extra neutral gauge bosons,''
  Phys.\ Rept.\  {\bf 317}, 143 (1999)
  [arXiv:hep-ph/9805494].
  %%CITATION = HEP-PH 9805494;%%

%\cite{Chay:1998hd}
\bibitem{Chay:1998hd}
  J.~Chay, K.~Y.~Lee and S.~h.~Nam,
  %``Bounds of the mass of Z' and the neutral mixing angles in general  SU(2)L x
  %SU(2)R x U(1) models,''
  Phys.\ Rev.\ D {\bf 61}, 035002 (2000)
  [arXiv:hep-ph/9809298].
  %%CITATION = HEP-PH 9809298;%%
%\cite{Jung:1999wq}
%\bibitem{Jung:1999wq}
  D.~W.~Jung, K.~Y.~Lee, H.~S.~Song and C.~Yu,
  %``Polarization effects on W boson pair productions with the extra neutral
  %gauge boson at the e+ e- linear collider,''
  J.\ Korean Phys.\ Soc.\  {\bf 36}, 258 (2000)
  [arXiv:hep-ph/9905353];
  %%CITATION = HEP-PH 9905353;%%
%\cite{Lee:2000km}
%\bibitem{Lee:2000km}
  K.~Y.~Lee, S.~C.~Park, H.~S.~Song and C.~Yu,
  %``Probing Z' gauge boson with the spin configuration of top quark pair
  %production at future e- e+ linear colliders,''
  Phys.\ Rev.\ D {\bf 63}, 094010 (2001)
  [arXiv:hep-ph/0011173].
  %%CITATION = HEP-PH 0011173;%%


\bibitem{Leroux:2001fx}
  K.~Leroux and D.~London,
  %``Flavour-changing neutral currents and leptophobic Z' gauge bosons,''
  Phys.\ Lett.\ B {\bf 526}, 97 (2002)
  [arXiv:hep-ph/0111246].
  %%CITATION = HEP-PH 0111246;%%


%\cite{Cvetic:1996mf}
\bibitem{Cvetic:1996mf}
  M.~Cvetic and P.~Langacker,
  %``New Gauge Bosons from String Models,''
  Mod.\ Phys.\ Lett.\ A {\bf 11}, 1247 (1996)
  [arXiv:hep-ph/9602424].
  %%CITATION = HEP-PH 9602424;%%


%\cite{Buchalla:1995dp}
\bibitem{Buchalla:1995dp}
  G.~Buchalla, G.~Burdman, C.~T.~Hill and D.~Kominis,
  %``GIM Violation and New Dynamics of the Third Generation,''
  Phys.\ Rev.\ D {\bf 53}, 5185 (1996)
  [arXiv:hep-ph/9510376];
  %%CITATION = HEP-PH 9510376;%%
%\cite{Burdman:2000in}
  G.~Burdman, K.~D.~Lane and T.~Rador,
  %``Anti-B B mixing constrains topcolor-assisted technicolor,''
  Phys.\ Lett.\ B {\bf 514}, 41 (2001)
  [arXiv:hep-ph/0012073].
  %%CITATION = HEP-PH 0012073;%%


%\cite{Cleaver:1998gc}
\bibitem{Cleaver:1998gc}
 G.~Cleaver, M.~Cvetic, J.~R.~Espinosa, L.~L.~Everett, P.~Langacker and J.~Wang,
  %``Physics implications of flat directions in free fermionic superstring
  %models. I: Mass spectrum and couplings,''
  Phys.\ Rev.\ D {\bf 59}, 055005 (1999)
  [arXiv:hep-ph/9807479];
  %%CITATION = HEP-PH 9807479;%%
%\cite{Cvetic:2001tj}
  M.~Cvetic, G.~Shiu and A.~M.~Uranga,
  %``Three-family supersymmetric standard like models from intersecting  brane
  %worlds,''
  Phys.\ Rev.\ Lett.\  {\bf 87}, 201801 (2001)
  [arXiv:hep-th/0107143];
  %%CITATION = HEP-TH 0107143;%%
%\cite{Cvetic:2001nr}
  M.~Cvetic, G.~Shiu and A.~M.~Uranga,
  %``Chiral four-dimensional N = 1 supersymmetric type IIA orientifolds from
  %intersecting D6-branes,''
  Nucl.\ Phys.\ B {\bf 615}, 3 (2001)
  [arXiv:hep-th/0107166];
  %%CITATION = HEP-TH 0107166;%%
%\cite{Cvetic:2002qa}
  M.~Cvetic, P.~Langacker and G.~Shiu,
  %``Phenomenology of a three-family standard-like string model,''
  Phys.\ Rev.\ D {\bf 66}, 066004 (2002)
  [arXiv:hep-ph/0205252].
  %%CITATION = HEP-PH 0205252;%%


%\cite{Masip:1999mk}
\bibitem{Masip:1999mk}
  M.~Masip and A.~Pomarol,
  %``Effects of SM Kaluza-Klein excitations on electroweak observables,''
  Phys.\ Rev.\ D {\bf 60}, 096005 (1999)
  [arXiv:hep-ph/9902467].
  %%CITATION = HEP-PH 9902467;%%


\bibitem{littlehiggs}
  N.~Arkani-Hamed, A.~G.~Cohen, E.~Katz, A.~E.~Nelson, T.~Gregoire and J.~G.~Wacker,
  %``The minimal moose for a little Higgs,''
  JHEP {\bf 0208}, 021 (2002)
  [arXiv:hep-ph/0206020];
  %%CITATION = HEP-PH 0206020;%%
  T.~Han, H.~E.~Logan, B.~McElrath and L.~T.~Wang,
  %``Phenomenology of the little Higgs model,''
  Phys.\ Rev.\ D {\bf 67}, 095004 (2003)
  [arXiv:hep-ph/0301040].
  %%CITATION = HEP-PH 0301040;%%

\bibitem{:1995fx}
  [ALEPH Collaboration],
  %``A Combination of preliminary LEP electroweak measurements and constraints
  %on the standard model,''
  CERN-PPE-95-172
  %\href{http://www.slac.stanford.edu/spires/find/hep/www?r=cern-ppe-95-172}{SPIRES entry}
  {\it Contributed to the 1995 Europhysics Conference on High Energy  Physics - EPS-HEP, Brussels, Belgium, 27 Jul - 2 Aug 1995,  and to the 17th
International Symposium on Lepton-Photon Interactions,  Beijing, China, 10 - 15 Aug 1995}

\bibitem{Abe:1996wy}
  F.~Abe {\it et al.}  [CDF Collaboration],
  %``Inclusive jet cross section in ${\bar p p}$ collisions at $\sqrt{s}=1.8$
  %TeV,''
  Phys.\ Rev.\ Lett.\  {\bf 77}, 438 (1996)
  [arXiv:hep-ex/9601008].
  %%CITATION = HEP-EX 9601008;%%

%\cite{Abbott:1997dr}
\bibitem{Abbott:1997dr}
  B.~Abbott {\it et al.}  [D0 Collaboration],
  %``Search for new particles decaying to two-jets with the D0 detector,''
FERMILAB-CONF-97-356-E
  %\href{http://www.slac.stanford.edu/spires/find/hep/www?r=fermilab-conf-97-356-e}{SPIRES entry}
  {\it Presented at 18th International Symposium on Lepton and Photon Interactions (LP 97), Hamburg, Germany, 28 Jul - 1 Aug 1997, and Presented at
  International Europhysics Conference on High-Energy Physics (HEP 97), Jerusalem, Israel, 19-26 Aug 1997}


\bibitem{Rizzo:1998ut}
T.~G.~Rizzo,
  %``Gauge kinetic mixing and leptophobic Z' in E(6) and SO(10),''
  Phys.\ Rev.\ D {\bf 59}, 015020 (1999)
  [arXiv:hep-ph/9806397].
  %%CITATION = HEP-PH 9806397;%%

\bibitem{Babu:1996vt}
  K.~S.~Babu, C.~F.~Kolda and J.~March-Russell,
  %``Leptophobic U(1)'s and the R_b - R_c Crisis,''
  Phys.\ Rev.\ D {\bf 54}, 4635 (1996)
  [arXiv:hep-ph/9603212];
  %%CITATION = HEP-PH 9603212;%%
  %``Implications of generalized Z Z' mixing,''
  {\it ibid.} {\bf 57}, 6788 (1998)
  [arXiv:hep-ph/9710441].
  %%CITATION = HEP-PH 9710441;%%

%\cite{Kim:2006}
\bibitem{Kim:2006}
  C.~S.~Kim, T.~Morozumi, Sechul~Oh and Chaehyun~Yu,
  %``Final state interactions and strong phases in B $\to$ D K, D pi
  %and B $\to$ pi pi decays,''
  work in progress.
  %%CITATION = HEP-PH ;%%

%\cite{Buchalla:1995vs}
\bibitem{Buchalla:1995vs}
  G.~Buchalla, A.~J.~Buras and M.~E.~Lautenbacher,
  %``Weak Decays Beyond Leading Logarithms,''
  Rev.\ Mod.\ Phys.\  {\bf 68}, 1125 (1996)
  [arXiv:hep-ph/9512380].
  %%CITATION = HEP-PH 9512380;%%

%\cite{Misiak:1999yg}
\bibitem{Misiak:1999yg}
  M.~Misiak and J.~Urban,
  %``{QCD} corrections to FCNC decays mediated by Z-penguins and W-boxes,''
  Phys.\ Lett.\ B {\bf 451}, 161 (1999)
  [arXiv:hep-ph/9901278];
  %%CITATION = HEP-PH 9901278;%%
  G.~Buchalla and A.~J.~Buras,
  %``The rare decays K $\to$ pi nu anti-nu, B $\to$ X nu anti-nu and  B $\to$ l+
  %l-: An update,''
  Nucl.\ Phys.\ B {\bf 548}, 309 (1999)
  [arXiv:hep-ph/9901288].
  %%CITATION = HEP-PH 9901288;%%



%\cite{Abe:2005bq}
\bibitem{Abe:2005bq}
  K.~Abe {\it et al.}  [Belle Collaboration],
  %``Search for B $\to$ tau nu and B $\to$ K nu anti-nu decays with a fully
  %reconstructed B at belle,''
  arXiv:hep-ex/0507034.
  %%CITATION = HEP-EX 0507034;%%

%\cite{Aubert:2004ws}
\bibitem{Aubert:2004ws}
  B.~Aubert {\it et al.}  [BABAR Collaboration],
  %``A search for the decay B+ $\to$ K+ nu anti-nu,''
  Phys.\ Rev.\ Lett.\  {\bf 94}, 101801 (2005)
  [arXiv:hep-ex/0411061].
  %%CITATION = HEP-EX 0411061;%%

%\cite{Ball:2004ye}
\bibitem{Ball:2004ye}
  P.~Ball and R.~Zwicky,
  %``New results on B $\to$ pi, K, eta decay formfactors from light-cone sum
  %rules,''
  Phys.\ Rev.\ D {\bf 71}, 014015 (2005)
  [arXiv:hep-ph/0406232].
  %%CITATION = HEP-PH 0406232;%%

%\cite{Ball:2004rg}
\bibitem{Ball:2004rg}
  P.~Ball and R.~Zwicky,
  %``B/(d,s) $\to$ rho, omega, K*, Phi decay form factors from light-cone sum
  %rules revisited,''
  Phys.\ Rev.\ D {\bf 71}, 014029 (2005)
  [arXiv:hep-ph/0412079].
  %%CITATION = HEP-PH 0412079;%%

%\cite{Eidelman:2004wy}
\bibitem{Eidelman:2004wy}
  S.~Eidelman {\it et al.}  [Particle Data Group],
  %``Review of particle physics,''
  Phys.\ Lett.\ B {\bf 592} (2004) 1.
  %%CITATION = PHLTA,B592,1;%%

%\cite{Langacker:1988ur}
\bibitem{Langacker:1988ur}
  P.~Langacker and D.~London,
  %``Mixing Between Ordinary And Exotic Fermions,''
  Phys.\ Rev.\ D {\bf 38}, 886 (1988).
  %%CITATION = PHRVA,D38,886;%%

%\cite{Asaka:2005pn}
\bibitem{Asaka:2005pn}
  T.~Asaka and M.~Shaposhnikov,
  %``The nuMSM, dark matter and baryon asymmetry of the universe,''
  Phys.\ Lett.\ B {\bf 620}, 17 (2005)
  [arXiv:hep-ph/0505013].
  %%CITATION = HEP-PH 0505013;%%

\end{thebibliography}
\end{document}